\documentclass[aps,prd,preprint,superscriptaddress,tightenlines,nofootinbib]{revtex4}

\long\def\symbolfootnote[#1]#2{\begingroup%
\def\thefootnote{\fnsymbol{footnote}}\footnote[#1]{#2}\endgroup}

\usepackage{amsfonts,amsmath,amsthm,amssymb}
\usepackage{mathrsfs}
\usepackage{bm}
\usepackage{color}
\usepackage{epsfig}

\newcommand{\PRE}[1]{{#1}}   % Use if preprint style

\newcommand{\beq}{\begin{equation}}
\newcommand{\eeq}{\end{equation}}
\newcommand{\bea}{\begin{flushleft} \begin{eqnarray}}
\newcommand{\eea}{\end{eqnarray}\end{flushleft}}

\newcommand{\comment}[1]{}

\newcommand{\ci}[1]{}
\newcommand{\ba}{\begin{eqnarray}}
\newcommand{\ea}{\end{eqnarray}}
\newcommand{\be}{\begin{equation}}
\newcommand{\ee}{\end{equation}}
\newcommand{\bay}[1]{\left(\begin{array}{#1}}
\newcommand{\eay}{\end{array}\right)}

\begin{document}

\preprint{
\hfil
\begin{minipage}[t]{3in}
\begin{flushright}
\vspace*{.4in}
MPP--2012--102\\
LMU-ASC 38/12\\
%CERN-PH-TH/2012-XXX\\
\end{flushright}
\end{minipage}
}

\title{\PRE{\vspace*{0.55in}}
Black Hole  Quantum Mechanics in the Presence of  Species
\PRE{\vspace*{0.5in}} }

\author{{\bf Gia Dvali}}\thanks{georgi.dvali@cern.ch}
\affiliation{Arnold Sommerfeld Center for Theoretical Physics 
Ludwig-Maximilians-Universit\"at M\"unchen,
80333 M\"unchen, Germany
\PRE{\vspace{.05in}}
}
\affiliation{Max--Planck--Institut f\"ur Physik,\\
Werner--Heisenberg--Institut,
80805 M\"unchen, Germany
\PRE{\vspace*{.05in}}
}
\affiliation{CERN,
Theory Department\\
1211 Geneva 23, Switzerland
\PRE{\vspace*{.05in}}
}
\affiliation{Center for Cosmology and Particle Physics\\
Department of Physics, New York University\\
4 Washington Place, New York, NY 10003, USA
\PRE{\vspace*{.05in}}
}

\author{{\bf Cesar Gomez}}\thanks{cesar.gomez@uam.es}

\affiliation{Arnold Sommerfeld Center for Theoretical Physics 
Ludwig-Maximilians-Universit\"at M\"unchen,
80333 M\"unchen, Germany
\PRE{\vspace{.05in}}
}

\affiliation{Instituto de F\'{\i}sica Te\'orica UAM-CSIC, C-XVI \\
Universidad Aut\'onoma de Madrid,
Cantoblanco, 28049 Madrid, Spain
\PRE{\vspace*{.05in}}}

\author{{\bf Dieter L\"ust}}\thanks{dieter.luest@lmu.de, luest@mppmu.mpg.de}
\affiliation{Arnold Sommerfeld Center for Theoretical Physics 
Ludwig-Maximilians-Universit\"at M\"unchen,
80333 M\"unchen, Germany
\PRE{\vspace{.05in}}
}
\affiliation{Max--Planck--Institut f\"ur Physik,\\
Werner--Heisenberg--Institut,
80805 M\"unchen, Germany
\PRE{\vspace*{.05in}}
}

\PRE{\vspace*{.25in}}

\begin{abstract}\vskip 2mm
 \noindent 
  Recently within the context of 
a microscopic quantum theory, the {\sl Black Hole's Quantum N-Portrait}, 
it was shown that continuous global symmetries are compatible with quantum black hole physics.
In the present paper we 
revise within the same framework the semi-classical black hole bound on the number 
of particle species $N_{species}$. We show that unlike the bound on global charge, the bound on species survives in the quantum picture and gives rise to a new fundamental length-scale, 
$L_{species} \, = \, \sqrt{N_{species}} \, L_P$, beyond which the resolution of species identities is impossible.   This finding nullifies 
the so-called species problem. This scale sets the size of 
the lightest quantum black hole in the theory, {\it Planckion}.  
 A crucial difference between the gravitational and non-gravitational 
 species emerges.   For gravitational species, the lightest black holes 
 are exactly at the scale of perturbative unitarity violation, which is a strong indication for self-UV-completion of gravity.  
 However,  non-gravitational species create a gap between the perturbative unitarity scale and 
 the lightest black holes, which must be filled by some unitarity-restoring physics. 
  Thus, self-UV-completion of gravity implies that the number of non-gravitational 
  species must not exceed the gravitational ones.  
\end{abstract}

\maketitle

\section{Introduction} 

 Semi-classically black holes are known to exhibit no-hair \cite{no-hair} and 
exact thermality of Hawking evaporation \cite{hawking}. 
 Despite the fact that these properties are derived in an idealized semi-classical limit in which black holes are infinitely heavy and infinitely long-leaved, one usually assumes that these are good 
(up to exponentially small deviations)  approximations for real black holes of finite mass and 
lifetime, as long as they are much heavier than the Planck mass ($M_P$).  

  Such extrapolation then leads to highly profound restrictions on the symmetry and the particle content 
  of the theory.  The two examples of such restrictions are: 1) The  'folk theorems" that state 
  incompatibility of exact global symmetries with gravity; and 2)  The restriction on the number of light particle 
  species, $N_{species}$ \cite{speciesbound}. 
  
   Both restrictions, when quantified lead to surprisingly similar constraints in the following sense. 
 Let us denote by $N_{global}$ the maximal possible dimensionality of a representation under a global symmetry group  allowed in a given theory.   Essentially $N_{global}$ counts the maximal amount of 
 the measurable global charge in a given theory. For continuous symmetries $N_{global}$ is infinite, since 
 a global charge (e.g., $U(1)$ baryon number) can be increased without a limit by considering the states with arbitrary large occupation number of the charged particles (e.g., baryons). 
 
  Let us denote by $N_{species}$ the number of particle species in the theory. For definiteness
  let us assume the species to be massless.  Of course $N_{global}$ and $N_{species}$ are 
  unrelated quantities.  One counts the number of species, whereas the other counts a maximal charge, 
  which can be arbitrarily large even if $N_{species}\,  = \, 1$. Despite this difference, in both cases, under the assumption of the validity of semi-classical 
 properties to finite-mass black holes, one can design thought experiments that lead to the emergence of very similar length-scales below which the theory must be modified.
    For the species case the scale is \cite{speciesbound}  
    \begin{equation}
      L_{species} \, = \, \sqrt{N_{species}}\, L_P \, .
   \label{lspecies}
   \end{equation}
 Whereas for global (or discrete gauge) symmetries the scale reads \cite{global}
  \begin{equation}
      L_{global} \, = \, \sqrt{N_{global}}\, L_P \, , 
   \label{lglobal}
   \end{equation}
  where  $ L_P$ is the Planck length that in terms of the Planck mass
 and Newton's constant can be expressed as
 $L_P \, \equiv \, {\hbar \over M_P} \,  \equiv \, \sqrt{ \hbar \, G_N }\,$.
 In the above semi-classical treatment,  both scales appear to have universal and  fundamental meaning, 
 since these are the shortest scales beyond which black hole no-hair and thermality properties must be fully violated.  Thus, both scales are treated as universal cutoff for applicability of Einstein gravity. 
 
 Both constraints would have far-reaching consequences.  For example, (\ref{lglobal})  immediately implies impossibility of having exact continuous global symmetries, since 
  in such a case one would have  $N_{global} \, = \, \infty$ which would imply $L_{global}\,  =\, \infty$.   
 
 However,  it is crucial to remember that in attributing a fundamental meaning to the scales $L_{global}$ and $L_{species}$, one is  based on the very strong assumption that semi-classical properties can be 
 extended to real (quantum) black holes of finite mass.   Whether this is a valid assumption can only be answered in a microscopic fully-quantum description. 
 
  This question was addressed in \cite{hair} 
 within the context of the recently-formulated quantum portrait of  black holes \cite{gia-cesar}.
  It was shown there that the extension of semi-classical no-hair arguments to real black holes 
  was incorrect, since black holes do carry hair even under global charges to the order $N_{global}/N$,  
  where $N$ is the occupation number of gravitons within the black hole, and $N_{global}$ is the amount of the global charge swallowed by the black hole. 
    As a result, the basic assumptions leading to  "folks theorems" about the non-existence of global charges are false.  Within the black hole quantum portrait,  the scale $L_{global}$ is fully 
  reproducible,  however it has no universal meaning.  It simply marks a stage in black hole's  evolution history  beyond which a particular black hole that initially swallowed $N_{global}$ units of a global charge depletes to the point where occupation numbers of gravitons and global charges equilibrates,  i.e.
$N_{global} \, \sim \, N$. Hence at this state,  the global  hair 
 becomes fully important.  At the same time a different black hole that was endowed by 
 a zero global charge,  will continue an almost-thermal depletion all the way till it reaches the size $\sim \, L_P$.

 So far the discussion was constrained to the pure gravity case. However particle species, which are coupled to gravitons,  exist in nature.  
 Therefore 
  in the present paper we shall extend the analysis of \cite{hair} applying it to the particle species hair.  In particular, we will  investigate the quantum meaning of the scale $L_{species}$.  We shall discover that unlike 
  $L_{global}$, the scale $L_{species}$ has an universal and a fundamental meaning. 
  It is a scale beyond which an {\it arbitrary}   black hole in a given theory, regardless of its initial composition, 
  ceases to deplete at an approximately-thermal rate.  Moreover, in full accordance with
  \cite{info},  $L_{species}$ is a lower cutoff beyond which the species identities cannot 
  be resolved.   
   As we shall see, the black hole quantum  portrait  reveals an important information about 
 the underlying  quantum  physics behind this scale, which is impossible to read-off in the standard semi-classical treatment. 
  Namely, the meaning of species scale is to set the size of lightest 
 quantum black holes existing in the theory. 
 
  Our findings have a number of important implications. 

  First, in full accordance with \cite{dvalisolodukhin} it nullifies the so-called species problem (for a review of species problem, see \cite{speciesproblem}). 
 One incarnation of this problem is a seeming inconsistency between 
 the Bekenstein and entanglement \cite{eentropy} entropies in the presence 
 of many particle species.   
  Our result confirms the claim of \cite{dvalisolodukhin},  that  this problem is an artifact of  not taking into the account the correct cutoff of the theory which 
is given by $L_{species}^{-1}$ rather than $M_P$.  
   
     Secondly,  our findings  have direct  implication for the concept of {\it self-UV-completion} of gravity by black hole physics \cite{self}.
     As explained in \cite{self} for self-completion of gravity it is essential that 
  the first quantum black holes exist right above the scale of perturbative unitarity violation. In this way, the black hole physics takes over and  theory can classicalize  exactly above the scale where 
  perturbative unitarity fails.
  
   As argued in \cite{self},  in pure Einstein gravity the two scales meet at $M_P$, and the 
existence of the lightest quantum black holes follows from the existence and the depletion of the heavier ones.  In the language of the quantum $N$-portrait
this claim becomes explicit in terms of occupation number of gravitons. 
The lightest quantum black holes are the ones with $N \, \sim \, 1$. 
Their mass is $M_P$, which coincides with perturbative 
scale of unitarity violation.   

  How the presence of species changes this balance? 
 As already suggested in \cite{self} the result must be sensitive to the nature of species.  If species are gravitational (such as e.g., Kaluza-Klein gravitons) the
 property of having lightest black holes exactly at the unitarity violation scale 
 does not change.  On the other hand non-gravitational species create a gap 
 between the two scales. 
 
  Our present analysis within the quantum portrait exactly confirms this picture. 
 We see, that gravitational species,  equally affect the depletion as well as 
 self-sustainability properties of the graviton condensate. As a result, 
 the lightest quantum black holes ($N\sim 1$) have the mass 
 $L_{species}^{-1}$, which as we shall show exactly coincides with the scale 
 of perturbative unitarity violation. 
  Thus, in this case, the black hole classicalization regime smoothly merges with the perturbative 
  unitarity regime without any gap. 
  
  In contrast,  when species are non-gravitational the balance is violated. 
  Non-gravitational species (similarly to gravitational ones) change the depletion law of the condensate, but fail to change the self-sustainability properties. 
   As a result, we get a finite  gap  of order  $\sim  N_{species}^{3/4}$ between 
 the scale of perturbative unitarity violation and the first quantum black hole. 
 
  Generalizing this result for arbitrary mix of gravitational and non-gravita\-tio\-nal species, we reach the following powerful conclusion. 
  
  $~~~$
  
  {\it The absence of a gap between the scale of perturbative unitarity violation and the lightest quantum black holes demands that the number of 
  non-gravitational species should not exceed the number on gravitational ones.}
 
 $~~~$
  
     Taking the idea of self-completion seriously, the existence of about 
   $100$ non-gravitational species in the Standard model can be taken as indication of at least equal number of gravitational ones below the scale 
 of $10^{17}$GeV.

 The paper is organized as follows. In the next section we will generalize the {\sl Black Holes's Quantum N-Portrait} including besides the gravitons also other particle species
 into the black hole bound state. As a result we will see that the species scale is a fundamental, universal length in the quantum picture. 
  We then give a physical meaning to this scale in terms of measurement 
of species identities.  Finally, we discuss the crucial difference between gravitational and non-gravitational species, and the connection with self-UV-completion of gravity. 
Throughout the paper in the equations we shall ignore the order one numerical factors,  
 which are unimportant for our analysis.

 \section{Species Hair} 
 
 \subsection{Quantum Meaning of the Species Scale}
     
 Before proceeding, let us set the stage by briefly formulating main aspects of the black hole quantum theory
 \cite{gia-cesar}.  For a detailed discussion the reader is referred to the original papers.  
   According to  this picture,  in pure gravity, i.e. in the absence of other species, the black holes represent a self-sustained bound-states (a Bose condensate) of  weakly interacting long-wavelength  gravitons. 
 The remarkable thing about this bound-state is that the condition of self-sustainability 
 can be satisfied for an arbitrary occupation number $N$ and fixes the characteristic wavelength of gravitons 
 as $R \, = \, \sqrt{N} L_P$. 
  All the other characteristics 
 of the system, such as the total mass ($M$) and the coupling of individual gravitons 
 ($\alpha$)
  are uniquely determined by $N$ as $\alpha \, =  \,  1/ N$ and total mass   
  $M\, = \,  \sqrt{N}  {\hbar \over L_P} $. \footnote{ Notice, that these are the typical large-$N$ relations in  't Hooft's sense \cite{tHooft}.}
 %  This similarity reveals an important feature that black holes 
%from quantum mechanical point of view are systems that obey rules of large-$N$ physics, 
%very similar to,  for example,  the picture of baryon \cite{witten}  in  QCD with large number of colors $N_c$.
%The analogy is that the role of number of colors $N_c$ which determined  the number of 
%baryon-constituents (quarks) is replaced by the number of gravitons $N$. 
%This also makes clear the crucial difference, unlike $N_c$ which is an input parameter of the theory 
%$N$ is a derived parameter that can take arbitrary values.  This is why baryons exist only for fixed 
%$N_c$ whereas the black holes exist for arbitrary $N$, and for arbitrary $N$ they are  {\it leaky}. 
%
  In this picture, we operate exclusively by quantum mechanical notions, such as 
 quantum coupling, wave-length and occupation number.   All the geometric and thermodynamical 
 characteristics, 
 such as horizon and temperature, are emergent as a result of semi-classical limit which corresponds 
 to double scaling limit  
   \begin{equation}  
   N \, \rightarrow \, \infty, ~~~L_P\, \rightarrow \, 0\,,   ~~~ R\equiv \sqrt{N}L_P \, =\, {\rm finite}, ~~~\hbar \, = \, {\rm finite}\, .
   \label{limit}
   \end{equation}
The notions of the geometric Schwarzschild radius $R$ and of the temperature $T \, = \, \hbar /R$ only emerge in this limit.  Thus,  the quantity $R$, which   quantum-mechanically is a characteristic wave-length of the condensate, 
 acquires a geometric meaning only in  the semi-classical limit. 
 
 Not surprisingly  many of the intrinsically-quantum properties (such as black hair)  become lost in this limit, 
 unless one carefully monitors their scaling also.  The above mentioned "folk theorems" about 
 violation of global charges are precisely an artifact of not taking into the account the existence and scaling 
 of the black hole hair \cite{hair}.  
 
% For example, it was shown\cite{hair} that black holes carry detectable hair  as $1/N$ effect per
 %information carrying quantum.  This fact is very important for understanding black hole information 
 %processing.  In particular, it  invalidates well-known  "folk theorems" stating that black holes global  
  
 Let us now assume that besides the graviton, the theory includes other massless species with total number of species 
being  $N_{species}$. Let us see how the existence of the extra species in the theory spectrum 
affects the depletion law. 
We start with a self-sustained condensate that is initially composed only out of $N$ gravitons. 
  In the absence of other species,  the leading contribution to the depletion rate is coming from the 
  process in which two graviton re-scatter in such a way that one of them gains the 
  above-escape energy and leaks out of the condensate.  The corresponding 
  Feynman diagram is depicted in Fig.1
  \vskip0.5cm
    \begin{figure}[ht]
    \vspace{-0.5cm}
\begin{center}
%\begin{picture}(185,235)(20,0)
%\put(-6,200){{$\rho$}}
%\put(110,-6){{$1+z$}}
\includegraphics[width=85mm,angle=0.]{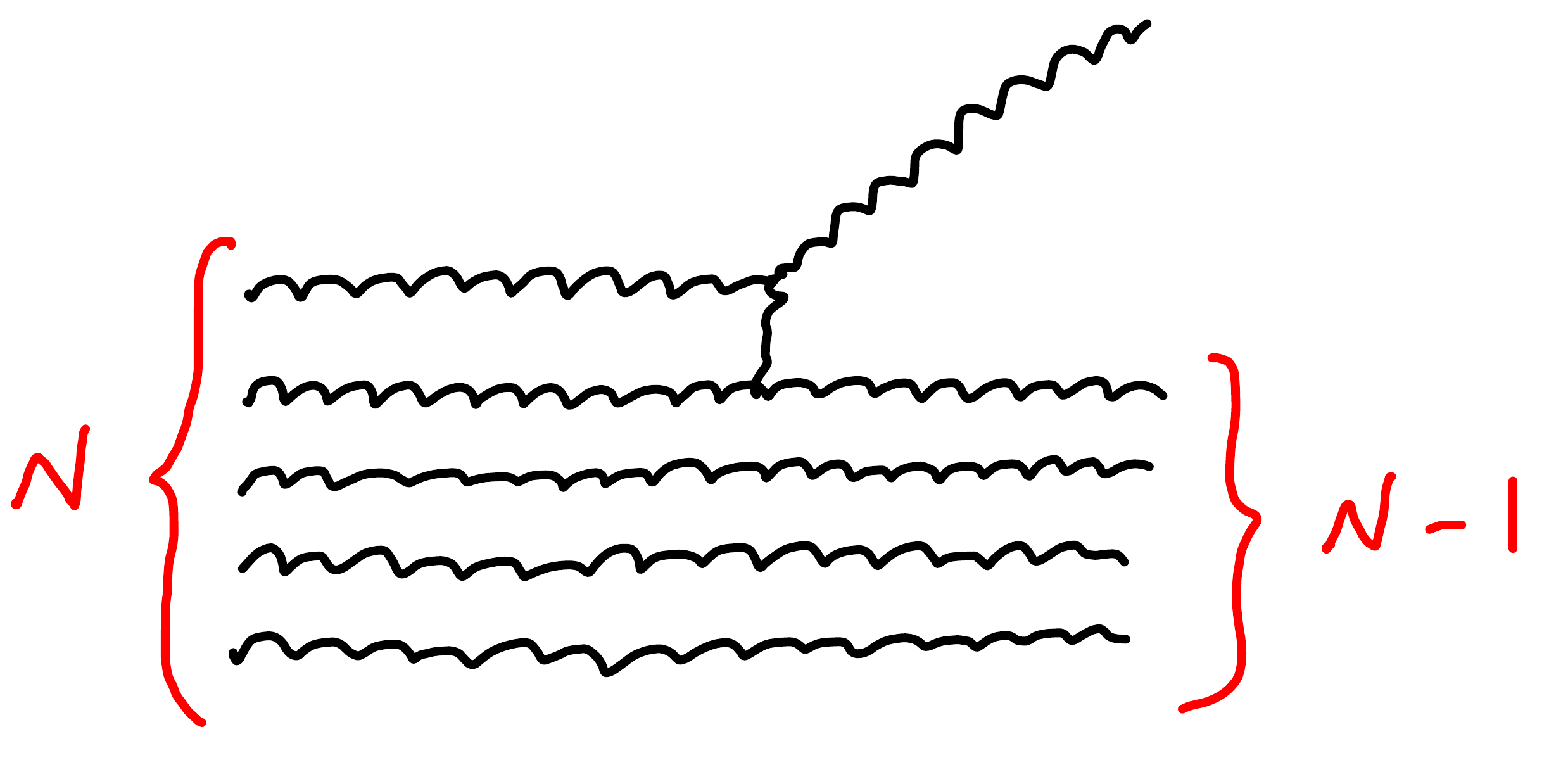}% Here is how to import EPS art
%\end{picture}
\end{center}
\caption{A leading order process responsible for quantum depletion of graviton condensate.
% In this process two out of  $N$ initial constituent gravitons scatter and one gains an above 
%escape energy.  
%  Initial $N$ gravitons  should not be considered as free, but as leading order 
%interaction eigenstates, with characteristic energy $\hbar /\sqrt{N}L_P$. 
}
\label{fig_safe}
\vspace{0.5cm}
\end{figure}

    The rate of this process 
  goes as, 
  \begin{equation}
 \Gamma_{leakage} \, = \,   {1 \over  \sqrt{N} L_P}   \, + \,  L_P^{-1} \,  {\mathcal O} (N^{-3/2}) \, ,  
\label{leakage}
\end{equation}
 which gives the following depletion equation,  
    \begin{equation}
   {\dot N} \, = -  {1 \over \sqrt{N} L_P}  \, + \, L_P^{-1} \,  {\mathcal O} (N^{-3/2}) \, . 
   \label{deplete}
   \end{equation}
  
    Notice that the quantity, which in the semiclassical limit acquires the meaning of Hawking temperature, is 
  \begin{equation}
  T \, \equiv  \, {\hbar  \over \sqrt{N}L_P} \, .
  \label{T}
  \end{equation}
  In terms of this parameter the depletion equation can be written as 
    \begin{equation}
   {\dot N} \, = -  {T \over  \hbar}  \, .  
   \label{depleteexact}
   \end{equation}
 Rewriting $N$ in terms of the black hole mass the depletion equation becomes a 
 Stefan-Boltzmann law for a black hole,
    \begin{equation}
   {\dot M} \, = -  {T^2 \over  \hbar}  \, . 
   \label{depletemass}
   \end{equation}
  Thus, the depletion imitates an approximately-thermal evaporation all the way until
  $N$ becomes order one, or equivalently, until the characteristic wave-length of the graviton condensate 
  becomes of order $L_P$. 
 
  Let us now see how the presence of extra species changes this situation.  
   In the presence of additional species, the contribution to the rate is enhanced
   by the processes, in which two gravitons now can annihilate also into the pair of two species. 
  % From this process
  % one gains the above-escape energy.   
   The corresponding diagram is  shown in Fig. 2.  
       \begin{figure}[ht]
       \vspace{-0.5cm}
\begin{center}
%\begin{picture}(185,235)(20,0)
%\put(-6,200){{$\rho$}}
%\put(110,-6){{$1+z$}}
\includegraphics[width=85mm,angle=0.]{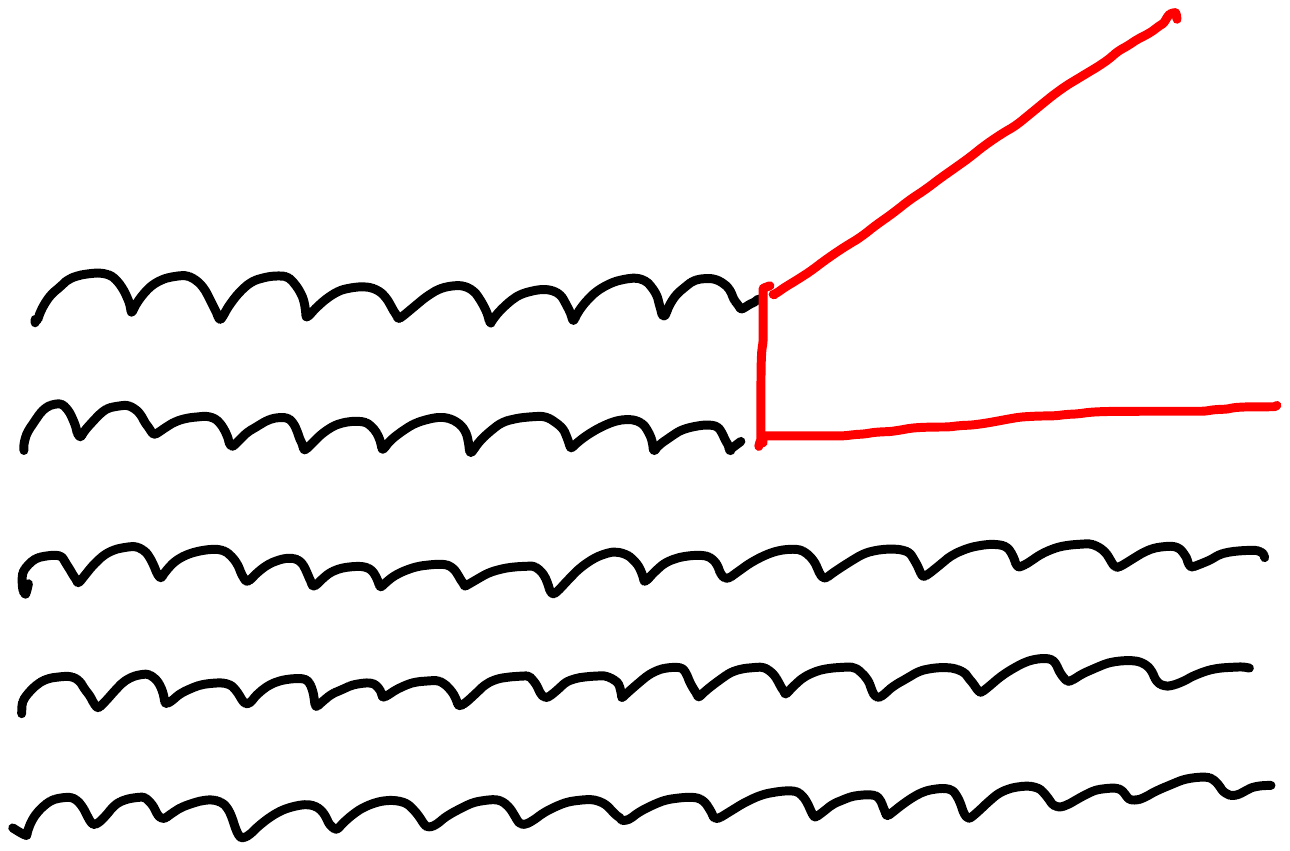}% Here is how to import EPS art
%\end{picture}
\end{center}
\vspace{-7cm}
\caption{A leading order process responsible for quantum depletion of graviton condensate into 
the species (represented by the red line).}
\label{fig_safe1}
\vspace{0.5cm}
\end{figure} 
  As a result the depletion rate is now enhanced by factor of $N_{species}$, 
    \begin{equation}
 \Gamma_{leakage} \, = \,   {N_{species} \over  \sqrt{N} L_P}   \, + \,  L_P^{-1} \,  {\mathcal O} (N^{-3/2}) \, ,  
\label{leakageNs}
\end{equation}
 which gives the following depletion equation,  
    \begin{equation}
   {\dot N} \, = -  {N_{species}  \over \sqrt{N} L_P}  \, + \, L_P^{-1} \,  {\mathcal O} (N^{-3/2}) \, . 
   \label{depleteNs}
   \end{equation}
   Notice,  that the quantum precursor of temperature is still given by the quantity (\ref{T}). 
   Taking the time-derivative of this equation and using (\ref{depleteNs}), we arrive at the following 
   relation, 
    \begin{equation}
    {{\dot T} \over T} \, = \, {N_{species} \over N}  \, T \, .
    \label{central}
    \end{equation}
     This equation tells us that as soon as the number of gravitons in the condensate depletes 
   to become comparable to $N_{species}$, the quantity  $T$ cannot even approximately be 
   interpreted as a temperature, since the rate of its change exceeds its own value.  
    The corresponding wavelength of the graviton condensate is given by, 
    \begin{equation}
      L_{species} \, = \, \sqrt{N_{species}}\, L_P \, .
   \label{lspecies}
   \end{equation}
  Thus, we  uncovered the quantum meaning of the species scale as the wave-length  
 of the self-sustained condensate, for which $N \, \sim \, N_{species}$. 
   This scale is an {\it universal} characteristics of a given theory. 
    Any black hole, irrespective of its initial composition, will loose even an approximate thermality
    ones its size will reach $L_{species}$. 
    
     We thus see that the scale $L_{species}$ maintains its universal meaning even in the quantum 
     picture.  It is also evident that this scale must have some fundamental meaning in the sense 
    that the black holes
    % (or into whatever states they crossover)
     must undergo a qualitative change in their nature  beyond this scale.   
    This is obvious from the fact that it makes no sense to talk about depleting self-sustained 
    bound-state of gravitons beyond this point, because the depletion time is much shorter than
   the wave-length.  
   
     %But,  how fundamental is this scale? 
         Moreover, let us show that we can never reach  purely gravitational bound-states of sizes smaller than $L_{species}$,
         even if we start out with a large black hole composed exclusively out of gravitons.          
 %        we start out with a large black hole it will get repopulated by  
 %   In order to  investigate this, we need to answer some other questions  first. 
  %   What happens with the depletion process beyond this point?        
 In order  to see this let us follow the evolution of a black hole which is initially 
 composed only out of gravitons.  
  The leading process of the depletion converts the black hole made out of $N$ gravitons into 
  the one with $N-2$ gravitons and one new species. So the black hole gets endowed   
 with species hair.   The probability of getting rid of this hair  is   suppressed by a factor $1/(N-2)$ relative to the
 probability of decreasing the number of gravitons.   So after every step the occupation number 
 of species in the black hole increases with the chance $N_{species}$ to one.   This process shall 
 continue until the number of other species and gravitons in the black hole equilibrates.
 Moreover, the numbers of species and their anti-particles must be roughly the same.  
 %This equilibration takes place for $N \, \sim \, N_{species}$.
 After this point the evolution of the condensate depends on the details of interaction between the species. 
 But, in any case the departure from thermality for such a condensate is maximal. 
   The study of  its dynamics is a very interesting problem per se, but beyond the scope of the present paper. 
    The lesson we would like to draw from this analysis is that it is highly unprobable that a
    purely graviton self-sustained bound-state can survive going through the scale 
  $L_{species}$.     
  % and gravitons.  
  %This reasoning shows that 
 %all black holes with $N \sim N_{species}$ irrespective of their initial composition will include 
 %same order of gravitons and other species in the condensate.  
 %Of course, the depletion of such a black hole 
 %can no longer be studied in $1/N$ expansion, but what is obvious is that it  is not even approximately thermal. 
   Thus the quantum portrait gives a simple microscopic rational  
 for the universal nature of the species scale $L_{species}$. 
 
  \subsection{Smallest Pixels of Nature}
       
    Finally, let us discuss another fundamental meaning of the species scale.  
  As suggested  in \cite{info} within the semi-classical treatment,  the species scale represents a lower bound on the scale down to which the species identities  can be resolved.  Let us see how this argument goes through the quantum portrait. 
  
  The semi-classical argument of \cite{info} is structured as follows. Imagine that we would like to build 
a  smallest pixel that could resolve the species identities in shortest possible time (that is, on the time-scale of its size).   
Let the size of the pixel be $L_{pixel}$. The necessary condition that such a pixel must satisfy is to 
 have localized sample of at least one member from each species (or an equivalent information).   Thus, we have at least  $N_{species}$ particles localized within the pixel.  The process of resolving the species identities is then a scattering process 
of an unknown particle at the pixel (see Fig. 3). 
    \begin{figure}[ht]
    \vspace{-0.5cm}
\begin{center}
%\begin{picture}(185,235)(20,0)
%\put(-6,200){{$\rho$}}
%\put(110,-6){{$1+z$}}
\includegraphics[width=75mm,angle=0.]{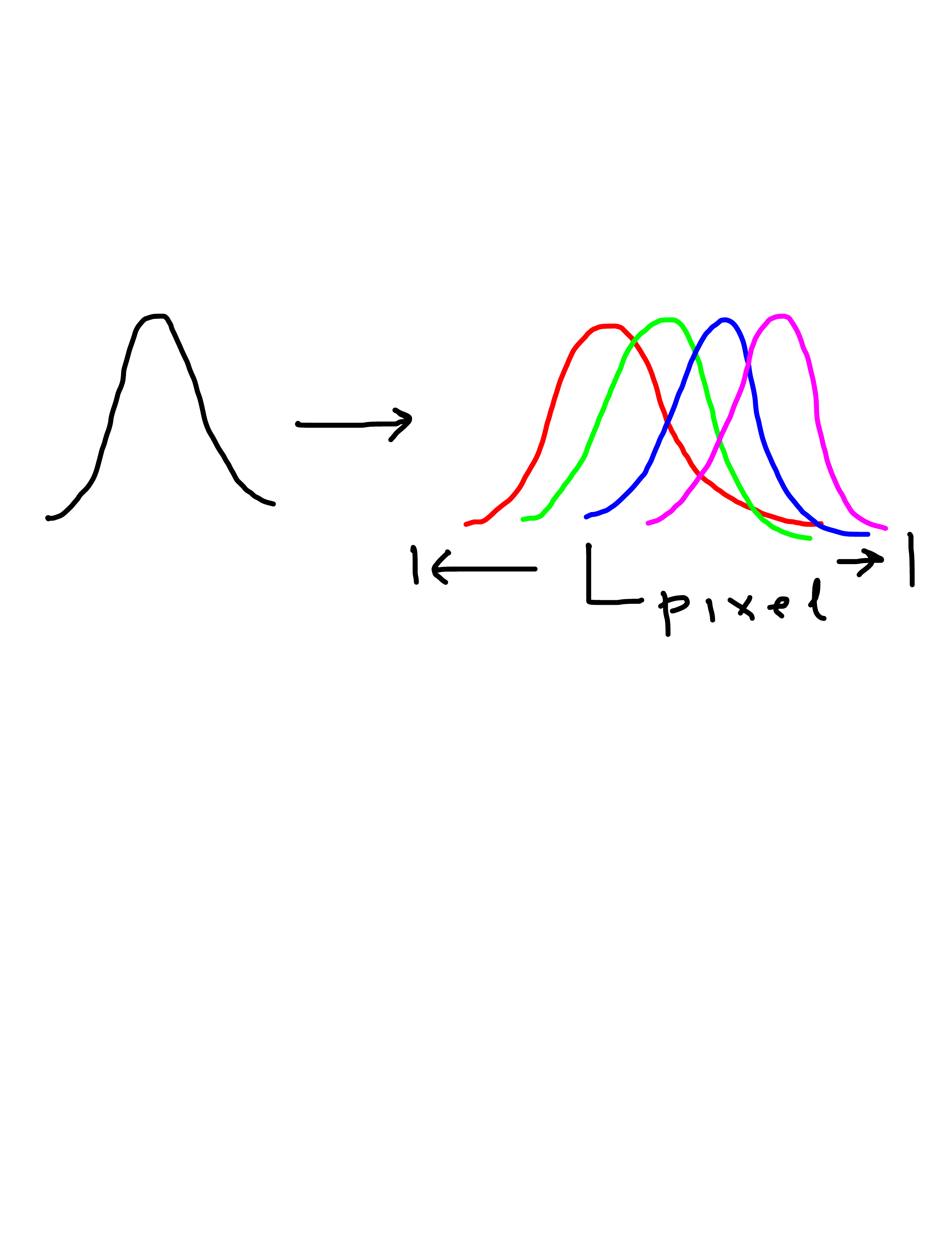}% Here is how to import EPS art
%\end{picture}
\end{center}
\vspace{-5cm}
\caption{The process of resolution of species identities. 
% in which a subject particle (represented in black)
%scatters off a pixel with localized different flavors of species represented in different colors.   
}
\label{fig_safe2}
%\vspace{0.5cm}
\end{figure}
 This requirement  gives the pixel a minimal mass, 
 \begin{equation}
  M_{pixel} \, = \,  N _{species} { \hbar \over L_{pixel}} \, .    
  \label{pixM}
  \end{equation}
  Then any attempt of decreasing the size of the pixel below $L_{species}$ would fail since it would result 
  into the  formation of a black hole of Schwarzschild radius exceeding $L_{species}$.  We thus arrive 
  to the lower cutoff being $L_{species}$.
  
  Let us now reinterpret this reasoning in our quantum language. 
  Quantum mechanically the pixel is a bound-state consisting of at least $N_{species}$ particles. 
 In reality, one needs an extra interaction to localize the species wave-functions within 
 $L_{pixel}$, which will increase the energy of the pixel beyond (\ref{pixM}); however  this only works in favor
 of our bound, since it increases the chances of black hole formation.  Now, according to the quantum 
 portrait \cite{gia-cesar},  the energy of localized species necessarily results into the occupation number of gravitons, given by $N \, = \, N _{species}^2 { L_P^2 \over L_{pixel}^2}$, which is convenient to rewrite as 
 \begin{equation}
  N \, = \, N_{species} \, {L_{species}^2 \over L_{pixel}^2}  \, .
  \label{gravitons}
  \end{equation}
   The latter form is very instructive, since it indicates that as soon as  we try to decrease $L_{pixel}$ below 
   $L_{species}$ the occupation number of gravitons will exceed the number of species, and moreover 
  the gravitons will form a self-sustained bound-state, a black hole with the occupation number of other species being a small fraction of the number of gravitons, $N_{species} \, \gg \, N$.
  The wavelength of the resulting graviton  bound-state (a size of a formed black hole) will be 
  \begin{equation}
   R \, = \, \sqrt{N} L_P \, \gg \, L_{species} 
   \label{pixelnew}
   \end{equation}
    Of course, this black hole carries a detectable species hair, but the probability of extracting 
  it is $1/N$ suppressed per species and thus takes much longer than the time $L_{species}$. 
   Namely, to detect a hair of each individual species takes the time \cite{hair}
   \begin{equation}
   t_{hair} \, = \, N^{3/2} L_P \, \gg \, L_{species} \, .
   \label{longtime}
   \end{equation}
  Thus, we see why the scale  $L_{species}$  is indeed a lower bound on the resolution capacity of the species identities.
  
   The above reasoning shows that the scale $L_{species}$ is fundamental even in quantum-mechanical sense.  In particular, 
  all the virtual processes in which species counts differently must be cutoff at this scale.
  Thus, in a theory with species the cutoff scale is $L_{species}$ and not $L_P$.  
   Necessity of such a cutoff immediately explains why there is no problem of in counting of the entanglement entropy in semi-classical limit
   \cite{dvalisolodukhin}.

 \section{Nature of Species}

  So far we were not interested in the nature of species.  We shall now become more specific and try to distinguish 
  between the {\it gravitational}  and {\it non-gravitational} species.  As we shall see,  the behavior of smallest black holes is dramatically different 
  in the two cases.  
 In case of gravitational species the mass of the lightest black holes  ("Planckions")  is right at the scale at which the perturbative violation of unitarity takes place.  
  As it was stressed in \cite{self}, this property is crucial for  
the self-completeness  of theory, since the unitarity is restored by quantum black holes.  

  In contrast, in case of non-gravitational species there is a mass gap between the scale of perturbative unitarity-violation and the first available quantum black holes.  UV-completion of theory demands that this gap be filled up by some 
  non-black hole physics. 
    
%  This difference suggests that for self-UV-completion of gravity  the number of non-gravitational species must be matched 
%  by gravitational ones.  Thus,  in the context of self-UV-completion the introduction of species implies change of  gravity (and thus emergence of new geometry) at short distances. 

 Let us now highlight this fundamental difference in more details. 
 
  \subsection{Gravitational Species} 
  
     We shell define {\it gravitational species} as the ones that just like graviton are sourced by the energy momentum of the existing particles in the theory. 
   Essentially, this set includes all the elementary  spin-$2$ species that may exist in the theory.  Of course, by general covariance all such states except one 
   (the "true" graviton) must be massive, and their number will obey a certain 
   distribution. 
 We shall parameterize this distribution  by denoting  $N_{species} (m)$ the number of all species  at and below a mass level $m$. A familiar example of such a distribution is 
 provided by Kaluza-Klein theories, although we shall keep our discussion 
 independent of geometric considerations. 
 
  The crucial property of gravitational species is that they not only affect the 
 depletion law, but also the self-sustainability condition of the condensate. 
 In order to make this clear,  we shall work in a particular example of {\it extreme  democratic} sourcing, in which all the species universally source each other 
 with the same strength as the massless graviton, $\alpha \, \equiv \, L_P^2/\lambda^2$, where $\lambda$ is a characteristic wave-length       
 in the process.  Of course,  in concrete cases the strength and the type of graviton couplings can (and will) obey certain selection rules, as it happens for example in Kaluza-Klein theories, however for our purposes this idealized approximation is good enough as it maximally sharpens the point.         
   
   In such a situation for the gravitons in a self-sustained bound-state of wave-length $\lambda$ the effective gravitational coupling becomes 
  \begin{equation}
    \alpha_{eff} \, = \, N_{species} (\lambda) \, \alpha \, .
 \label{effective}
 \end{equation}
  This fixes the self-sustainability relation between the wave-length and the occupation number $N$ as, 
\begin{equation}
  \lambda \, = \, \sqrt{N N_{species}} L_P \, = \, \sqrt{N} L_{species} \, ,  
  \label{wave}
  \end{equation}
  where the  $L_{species} \, = \, L_P \sqrt{N_{species}(\lambda)}$ has to be understood as  "running" species scale due to the wave-length dependence 
  of the number of species involved. 
  
   The depletion rate in such a case is 
   \begin{equation}
   \Gamma \, = \, \lambda^{-1} \, (L_p^2 /\lambda^2)^2 \, N^2 \, N_{species}^2 \, .
  \label{rategraviton}
   \end{equation}
   where the $N_{species}^2$ factor comes from the fact that, due to 
  democratic inter-sourcing, each constituent pair can annihilate into    
  $N_{species}^2$ pairs, as it is depicted on Fig. 4. 
  
      \begin{figure}[ht]
    \vspace{-1.5cm}
\begin{center}
%\begin{picture}(185,235)(20,0)
%\put(-6,200){{$\rho$}}
%\put(110,-6){{$1+z$}}
\includegraphics[width=105mm,angle=0.]{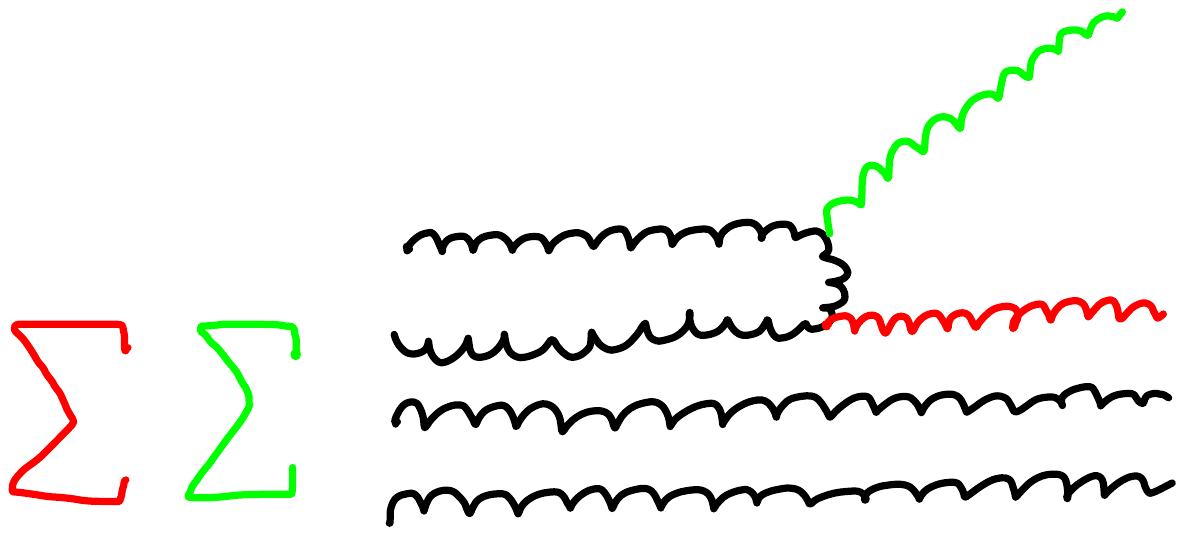}% Here is how to import EPS art
%\end{picture}
\end{center}
\vspace{-9.8cm}
\caption{A leading order process responsible for quantum depletion of graviton condensate in the presence of gravitational species. The summation goes over 
pairs of final states represented by green and red lines. 
% In this process two out of  $N$ initial constituent gravitons scatter and one gains an above 
%escape energy.  
%  Initial $N$ gravitons  should not be considered as free, but as leading order 
%interaction eigenstates, with characteristic energy $\hbar /\sqrt{N}L_P$. 
}
\label{fig_safe}
%\vspace{-0.5cm}
\end{figure}

  \vspace{0.5cm}

   Taking into the account  
 (\ref{wave}), the above rate can be rewritten is a very simple form,   
     \begin{equation}
   \Gamma \, =
   \, {1 \over  \sqrt{N} L_{species}} \, .
  \label{rateeasy}
   \end{equation}
   Comparing with (\ref{leakage}) we see that depletion law  is the same as 
   for a single species, but with $L_P$ being replaced by $L_{species}$. 
   
    This is an extremely important fact, which tells us that for gravitational species 
    the depletion law remains approximately thermal all the way till $N\sim 1$. 
 Notice, that the state $N\sim 1$,  is a lowest mass member of the black hole tower.   
    
    The mass and the characteristic wavelength of such a lightest  black hole is, 
  \begin{equation}
    M_{min} \, = \, L_{species}^{-1}, ~~~ \lambda_{min} \, = \, L_{species} \, .
  \label{mini}
  \end{equation}
  Notice that this coincides with the scale of perturbative unitarity violation in the theory.  Indeed, consider a tree-level $2\rightarrow 2$ scattering process 
depicted in Fig. 5, in which  a two initial particles annihilate  into any of the 
two species. 

    \begin{figure}[ht]
    \vspace{-1.5cm}
\begin{center}
%\begin{picture}(185,235)(20,0)
%\put(-6,200){{$\rho$}}
%\put(110,-6){{$1+z$}}
\includegraphics[width=115mm,angle=0.]{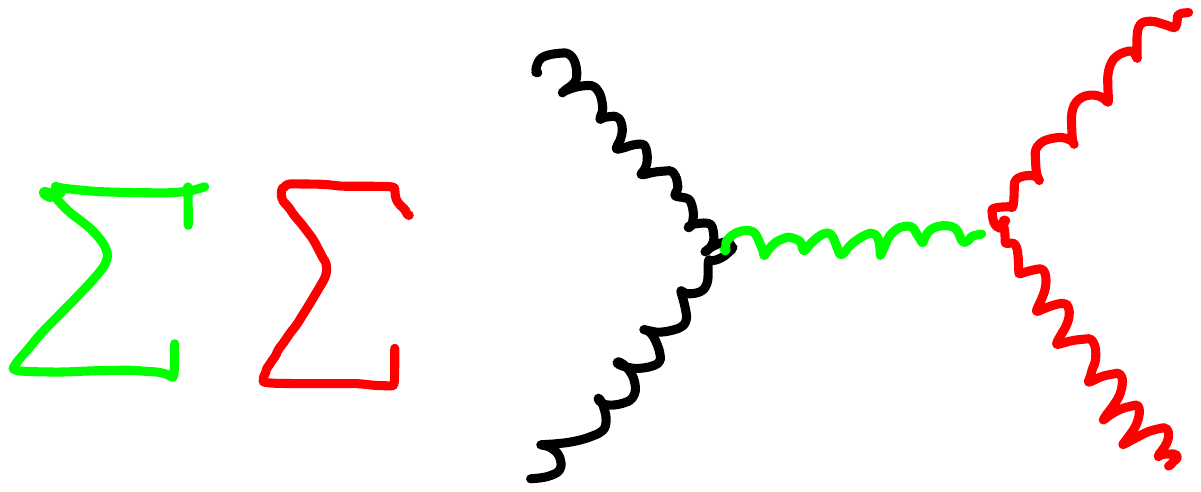}% Here is how to import EPS art
%\end{pictursce}
\end{center}
\vspace{-10.5cm}
\caption{A  $2\rightarrow 2$ scattering process of gravitational species 
responsible for violation of perturbative unitarity above $L_{species}^{-1}$.
% In this process two out of  $N$ initial constituent gravitons scatter and one gains an above 
%escape energy.  
%  Initial $N$ gravitons  should not be considered as free, but as leading order 
%interaction eigenstates, with characteristic energy $\hbar /\sqrt{N}L_P$. 
}
\label{fig_safe}
\vspace{0.5cm}
\end{figure}

  At center of mass energy  $E$, the rate of the process goes as 
\begin{equation}
  \Gamma_{2 \rightarrow 2} \, = \, E \, {E^4 \over M_P^4} \, N_{species}^2 
  \, = \, E (E L_{species})^4 \, ,
\end{equation} 
where the $N_{species}^2$ factor appears due to summation over 
both internal (green) and external (red) lines.  
 Obviously, this process violates perturbative unitarity at the energy 
\begin{equation}
 M_{unitary} \, = \, L_{species}^{-1}  
 \label{Munitarity1}  
 \end{equation}    
        
  Comparing with (\ref{mini}) we see that the quantum black holes enter into the game right above the unitarity violating scale $L_{species}$.
  The picture  is essentially the same as for single species, with the only difference that everything happens at a lower energy scale, with $L_{species}$  which  takes up the role of $L_P$.
   Thus, the necessary condition for self-completion \cite{self} is satisfied. The quantum 
   black holes are readily available in the spectrum right above the unitarity scale!
 
 \subsection{Non-Gravitational Species} 
   
   It is obvious that situation is very different  for non-gravitational species. 
  Such species affect the depletion law, but at the same time do not make 
  binding force stronger. As a result, the black hole goes out of thermal regime 
  while still being a multi-particle state,  
  much heavier than the perturbative-unitarity violation scale.  Indeed, since 
  non-gravitational species only affect the depletion law and not 
  the self-sustainability condition of the graviton condensate.  
  \footnote{This should not to be confused with the perturbative loop contribution to the 
  Newton's constant.  This contribution exists for both gravitational and non-gravitational species and it is simply summed up in fixing the effective Newton's constant once and for all.}  
  The  mass 
  of the smallest black hole in such a case is $M_{min} \, = \, L_{species} \, M_P^2 \, 
  = \, \sqrt{N_{species}} M_P$. 
  Notice that this  exceeds the perturbative unitarity violating scale, which for 
  non-gravitational species is at least as low as, 
  \begin{equation}  
  M_{unitary} \, = \,  M_P/N_{species}^{1/4}   \, .
  \label{Munitary2}
  \end{equation} 
 This unitarity bound comes from a $2\rightarrow 2$ scattering process 
 between the species via graviton exchange, depicted in Fig.6. 
The rate of this process goes similar to eq.(22), except there is one less power of $N_{species}$ 
due to the fact that there is only one intermediate graviton exchange. 
 
 \vspace{2cm}
 
     \begin{figure}[ht]
    \vspace{-3.0cm}
\begin{center}
%\begin{picture}(185,235)(20,0)
%\put(-6,200){{$\rho$}}
%\put(110,-6){{$1+z$}}
\includegraphics[width=105mm,angle=0.]{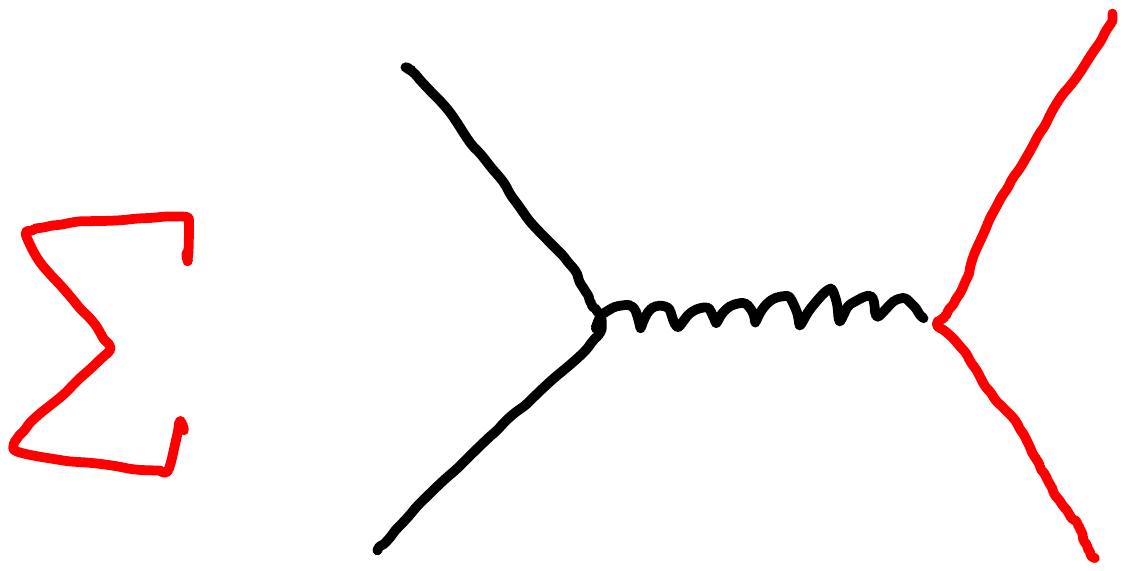}% Here is how to import EPS art
%\end{picture}
\end{center}
\vspace{-10cm}
\caption{A  process giving a perturbative  unitarity bound in which two 
initial particles are scattered into two final species represented by red lines. 
% In this process two out of  $N$ initial constituent gravitons scatter and one gains an above 
%escape energy.  
%  Initial $N$ gravitons  should not be considered as free, but as leading order 
%interaction eigenstates, with characteristic energy $\hbar /\sqrt{N}L_P$. 
}
\label{fig_safe}
\vspace{0.5cm}
\end{figure}

  So the question is what fills the unitarity gap between this scale and the smallest black holes? One way or another, we need some non-black hole 
  physics for curing the unitarity problem in the interval of energies 
 \begin{equation}
 M_P/N_{species}^{1/4} \, < \, E \,  < \sqrt{N_{species}} M_P \, .
 \label{window}
 \end{equation}  
  This situation is schematically depicted in Fig 7. 
  The red lines represented regions of perturbative unitarity, whereas the green ones the regions in which unitarity is restored by quantum black holes. The dashed region in the left is a non-unitarity   
window in case of non-gravitational species. 

$~~~$
  
    \begin{figure}[ht]
    \vspace{-1.5cm}
\begin{center}
%\begin{picture}(185,235)(20,0)
%\put(-6,200){{$\rho$}}
%\put(110,-6){{$1+z$}}
\includegraphics[width=65mm,angle=0.]{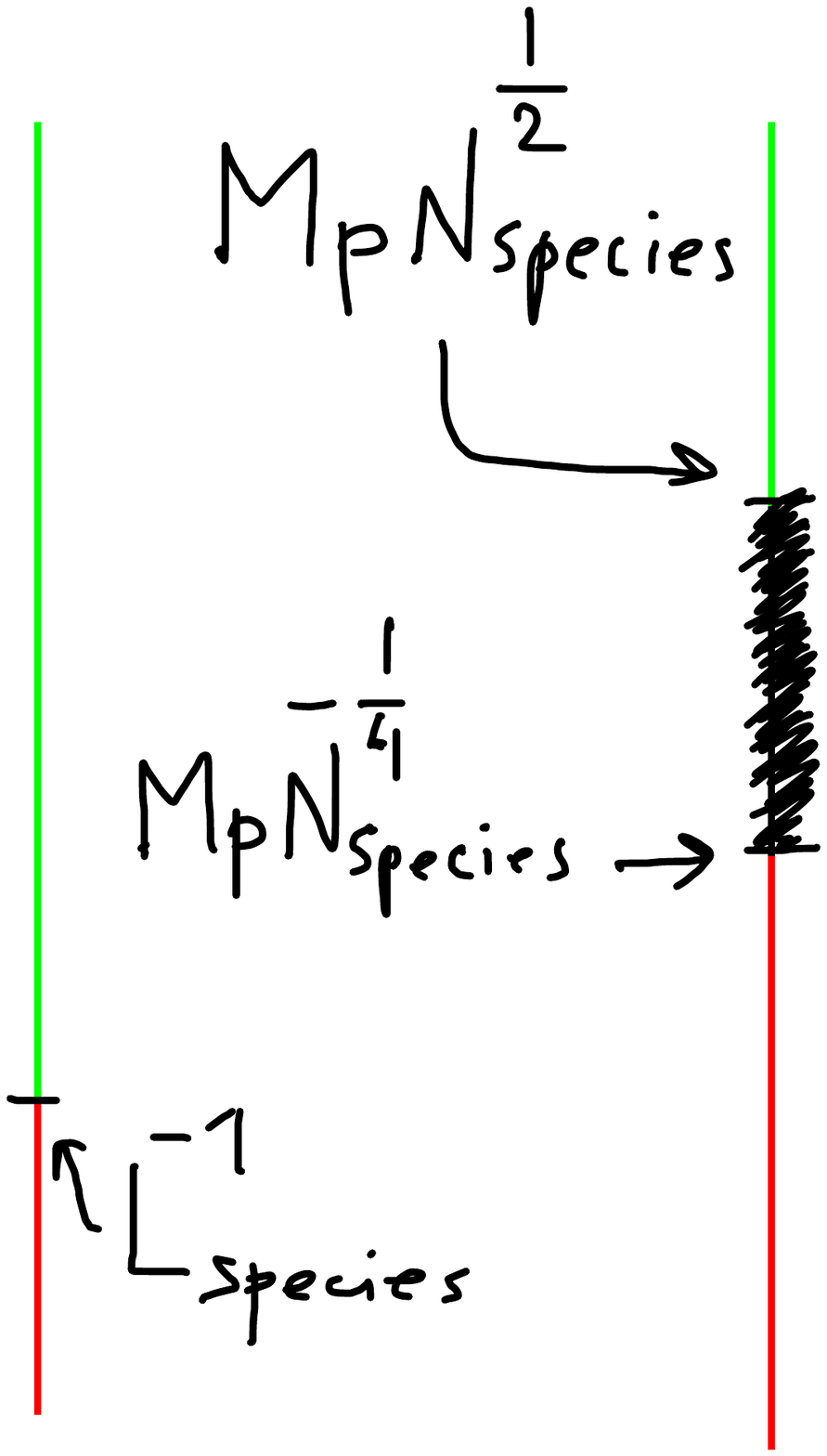}% Here is how to import EPS art
%\end{picture}
\end{center}
\vspace{-3cm}
\caption{ A schematic confrontation between the cases of gravitational (left) versus non-gravitational (right) species.  The red and green lines
represent the pertubative unitarity and black hole regions respectively. 
The dashed black region on the right represents a non-unitarity window in 
case of non-gravitational species.    
%The red lines represented regions of perturbative unitarity, whereas the green ones the regions in which in which unitarity is restored by quantum black holes. The dashed region in the left is a non-unitarity   
%window in case of non-gravitational species. 
}
\label{fig_safe}
\vspace{0.5cm}
\end{figure}

  Our analysis can be easily generalized to a mixed situation when species consist of 
  both gravitational and non-gravitational ones. 
 The particular details of course change, but the following conclusion 
 persists. Whenever number of non-gravitational species is much 
 larger than the gravitational ones, there is gap between the unitarity scale and 
 the masses of the smallest black holes. 
 
  This suggest the following surprisingly powerful conclusion: 
  
  {\it Self-UV-completion of gravity demands that the number of 
 non-gravitational species be less or equal than the gravitational ones 
 (normalized per strength of Einstein's graviton)}.

    \subsection{Discussions and Outlook} 
  
    In this paper we have reconsidered the semi-classical black hole bound on the 
 number of particle species \cite{speciesbound}  within the microscopic quantum portrait and have 
 shown that the bound persists in full quantum picture. 
  
   Introduction of $N_{species}$ in the theory creates a new fundamental length 
  scale, $L_{species}$,  which is an ultimate lowest cutoff on the black hole size.  
   Within quantum portrait this scale acquires a very clear physical meaning, as 
   the characteristic black hole size (or to be more precise, the characteristic wave-length of the black hole constituents) beyond which the depletion process {\it not even approximately} can mimic the Hawking thermal spectrum.  Essentially,  no 
   black holes of the smaller size exist in a given theory. 
   
    Furthermore,  we have shown that physics of the cutoff is highly sensitive 
    to the nature of species. 
  For gravitational species, the scale $L_{species}^{-1}$ simultaneously marks 
  the upper bound on the scale of perturbative unitarity, as well as the lowest bound on the black hole mass.  This is a strong indication in favor of self-UV-completion of gravity by classicalization through quantum black holes.  
  Indeed, the quantum black holes that are crucial for unitarizing scattering amplitudes are readily available right at the scale of perturbative unitarity violation.  
  
      Situation is dramatically different for non-gravitational species (such as 
      e.g., extra neutrino species). In this case there is a mass gap between the  
  mass of the lightest black holes and the scale of perturbative unitarity. 
   This gap, unless filled by some yet unknown physics, indicates an intrinsic inconsistency of the theory. 
   
    We are thus, lead to the conclusion that in any consistent theory of gravity 
    the number of gravitational species must be the dominant one. 
  All the known examples of deformation of Einstein gravity, such as Kaluza-Klein 
  or perturbative string theory,  satisfy this property.

\section*{Acknowledgements}

The work of G.D. was supported in part by Humboldt Foundation under Alexander von Humboldt Professorship,  by European Commission  under 
the ERC advanced grant 226371,   by TRR 33 \textquotedblleft The Dark
Universe\textquotedblright\   and  by the NSF grant PHY-0758032. 
The work of C.G. was supported in part by Humboldt Foundation and by Grants: FPA 2009-07908, CPAN (CSD2007-00042) and HEPHACOS P-ESP00346.
This work of D.L. is supported by  the
DFG Transregional Collaborative Research Centre TRR 33
and the DFG cluster of excellence `Origin and Structure of the Universe'.


\begin{thebibliography}{99}
    
    
    \bibitem{no-hair} 

  W.~Israel,  {\it Phys. Rev.} {\bf 164} (1967) 1776;  {\it Commun. Math. Phys.}
{\bf 8}, (1968) 245;  

B.~ Carter, {\it Phys. Rev. Lett.}  {\bf 26}  (1971)  331.

J.~Hartle, {\it Phys. Rev.}  {\bf D 3} (1971) 2938.

  J.~Bekenstein, {\it Phys. Rev. }\  {\bf D 5}, 1239 (1972);
  {\it Phys.\ Rev.}\  {\bf  D 5},  (1972) 2403; {\it Phys. Rev. Lett.} {\bf 28} (1972) 452. 
  
  C.~Teitelboim, {\it Phys. Rev.} {\bf D 5} (1972) 294.  
 

%\bibitem{paradox} 
%S. W. Hawking, Phys. Rev. D14 (1976) 2460
    
\bibitem{hawking}
S. W. Hawking, Comm. Math. Phys. 43 (1975) 199

%\bibitem{Bekenstein}

%J. D. Bekenstein, Phys. Rev. D7 (1973) 2333

    
    \bibitem{speciesbound} 

G.~Dvali, ``Black Holes and Large N Species Solution to the
Hierarchy Problem,'' arXiv:0706.2050 [hep-th];
G.~Dvali and M.~Redi, ``Black Hole Bound on the Number of Species and Quantum Gravity at LHC,''
Phys. Rev.  {\bf D77} ( 2008) 045027,  arXiv:0710.4344 [hep-th].

\bibitem{global} G. Dvali, M. Redi, S. Sibiryakov, A. Vainshtein, 
Gravity Cutoff in Theories with Large Discrete Symmetries.
 Phys.Rev.Lett. 101 (2008) 151603;   arXiv:0804.0769 [hep-th].

 Similar reasoning, but without derivation of the bound was given in,  

T.~Banks,  N.~ Seiberg, 
Symmetries and Strings in Field Theory and Gravity, 
 Phys.Rev. D83 (2011) 084019; arXiv:1011.5120 [hep-th]. 


%\cite{Dvali:2012rt}
\bibitem{hair}
  G.~Dvali and C.~Gomez,
  ``Black Hole's 1/N Hair,''
  arXiv:1203.6575 [hep-th].
  %%CITATION = ARXIV:1203.6575;%%
    
 \bibitem{gia-cesar}    
 	G.~Dvali, C.~Gomez,  Black Hole's Quantum N-Portrait, 
 arXiv:1112.3359 [hep-th]; 
 Landau-Ginzburg Limit of Black Hole's Quantum Portrait: Self Similarity and Critical Exponent,  arXiv:1203.3372 [hep-th]. 
 
 
\bibitem{info} 
G.~Dvali and C~Gomez, ``Quantum Information and Gravity Cutoff in Theories with Species'',
Phys. Lett. {\bf B674}  (2009) 303,  arXiv:0812.1940 [hep-th] . 


   


  

 %\cite{Dvali:2008jb}
\bibitem{dvalisolodukhin}
  G.~Dvali and S.~N.~Solodukhin,
  ``Black Hole Entropy and Gravity Cutoff,''
  arXiv:0806.3976 [hep-th].
  %%CITATION = ARXIV:0806.3976;%%
  
  \bibitem{speciesproblem}
 J. D. Bekenstein, Do we understand black hole entropy?, arXiv:gr-qc/9409015.

 
 \bibitem{eentropy}
 W. Israel, Phys. Lett. A57 (1976) 107;
L. Bombelli, R. K. Koul, J. H. Lee and R. D. Sorkin, Phys. Rev. D 34, 373 (1986);
M. Srednicki, Phys. Rev. Lett. 71, 666 (1993);
V. P. Frolov and I. Novikov, Phys. Rev. D 48, 4545 (1993).
 

  
  

\bibitem{self}
G.~ Dvali  and C.~ Gomez, Self-Completeness of Einstein Gravity,  arXiv:1005.3497 [hep-th].


\bibitem{tHooft} G.~'t Hooft, A Planar Diagram Theory for Strong Interactions.
 Nucl.Phys. B72 (1974) 461.


%\bibitem{class}
%G.~ Dvali, G.~ F. Giudice, C.~ Gomez, A.~ Kehagias,  UV-Completion by Classicalization, arXiv:1010.1415 [hep-ph]. JHEP 2011 (2011) 108 


%\bibitem{master}
%E. Witten, in Recent Developments in Gauge Theories eds. G. 't Hooft et. al. Plenum Press, New York and London (1980).






%\bibitem{Page} 
%D.N.~ Page, Information in black hole radiation.  Phys.Rev.Lett. 71 (1993) 3743-3746;  
%hep-th/9306083




   
\end{thebibliography}
\end{document}